# Training-Free Color-Aware Adversarial Diffusion Sanitization for Diffusion Stegomalware Defense at Security Gateways

Vladimir Frants, Sos Agaian

## Abstract

The rapid expansion of generative AI has normalized large-scale synthetic media creation, enabling new forms of covert communication. Recent generative steganography methods, particularly those based on diffusion models, can embed high-capacity payloads without fine-tuning or auxiliary decoders, creating significant challenges for detection and remediation. Coverless diffusion-based techniques are difficult to counter because they generate image carriers directly from secret data, enabling attackers to deliver stegomalware for command-and-control, payload staging, and data exfiltration while bypassing detectors that rely on cover–stego discrepancies. This work introduces Adversarial Diffusion Sanitization (ADS), a training-free defense for security gateways that neutralizes hidden payloads rather than detecting them. ADS employs an off-the-shelf pretrained denoiser as a differentiable proxy for diffusion-based decoders and incorporates a color-aware, quaternion-coupled update rule to reduce artifacts under strict distortion limits. Under a practical threat model and in evaluation against the state-of-the-art diffusion steganography method Pulsar, ADS drives decoder success rates to near zero with minimal perceptual impact. Results demonstrate that ADS provides a favorable security-utility trade-off compared to standard content transformations, offering an effective mitigation strategy against diffusion-driven steganography.

## 1. Introduction

Cyberattacks are increasing as offenders earn substantial profits with minimal risk, unlike traditional crimes. The worldwide economic toll of cybercrime is estimated to reach 6 trillion US dollars each year 2021 [1]. Steganography hides information inside benign-looking media and can bypass monitoring and content inspection controls [2], [3]. As network defenses improve, cybercriminals are more frequently using steganographic techniques.

This has led to a new form of malware called stegomalware, which uses steganographic techniques and has attracted researchers' interest. Stegomalware is designed to conceal malicious code, communications, or stolen data within seemingly ordinary files or network traffic, making them difficult to detect with standard security tools. Identifying stegomalware is challenging because it (i) hides command-and-control (C2) traffic and secondary malware payloads, making post-infection activities hard to detect [3]; and (ii) uses techniques like LSB and pattern-based methods across media such as images, text, and network data, often encrypting payloads to enhance concealment.

In security deployments, this enables real attacks: payloads, commands, and keys can be embedded in images to support staged malware delivery and covert command-and-control (C2), which reduces the value of signature-based scanning and format validation alone [4]. Since images are routinely exchanged across social platforms, messaging applications, enterprise email, app ecosystems, and Content Distribution Networks (CDNs), security gateways need defenses that preserve image utility for legitimate use while preventing them from serving as covert communication channels.

Classical image steganography typically embeds messages by modifying an existing cover image, which can introduce artifacts that steganalysis can exploit under suitable assumptions [5], [6]. Also, the widespread use of generative artificial intelligence (AI) has transformed the creation and distribution of digital media. Early systems of generative steganography based on GANs and Flow models demonstrated high data capacity and greater resistance to steganalysis. However, they often compromised realism due to training instability, mode collapse, or imperfect distribution matching. Current models, such as diffusion models and generative adversarial networks (GANs), routinely produce photorealistic images at scale, making synthetic content a common part of both consumer and enterprise workflows. This method may mitigate many of the classical issues in image steganography by leveraging robust score-based sampling and well-calibrated denoising.

Generative steganography involves embedding secret messages during image creation, rather than within pre-existing cover images. Different diffusion techniques can hide payloads, such as malicious code or command data, without requiring fine-tuning of the base generative model or creating extra extractors or decoders.

This approach challenges traditional methods that identify differences or artifacts between cover and stego images after they are created. A common defensive tactic is detection-first: classifying images as stego or non-stego and applying countermeasures only to those flagged. In diffusion-based coverless steganography, this strategy is vulnerable for three practical reasons. First, generating images without a cover eliminates the "reference cover" that many classical assumptions depend on [5], [6], and diffusion-based methods note that creating stego images without a cover reduces their vulnerability to steganalyzers because "there are no cover images for the steganalyzer to train on" [7].

Second, several designs aim to maintain the model's appearance by limiting message-induced changes, so that latent or noise variables remain close to the Gaussian prior, thereby weakening distributional tests and complicating detection by general-purpose detectors [7], [8]. Third, hiding schemes are rapidly evolving and can be fine-tuned against known detection methods, creating a high maintenance burden for fixed detectors in large-scale operations; this is especially evident in adversarial setups and continues in diffusion-based techniques that vary the embedding location through sampling, inversion, conditioning, and prompt/key mechanisms [7], [9]–[11]. While these innovations offer new opportunities, they also create covert channels that are difficult to detect and counter. Therefore, security systems that perform large-scale image processing should not rely solely on steganalysis for complete accuracy.

This work emphasizes proactive sanitization at the security gateway, which treats each incoming image as untrusted. It applies a transformation designed to maintain user-perceived utility while making diffusion-based payload recovery unreliable. This approach aligns with content disarm and reconstruction (CDR), where inputs are converted into safer formats before downstream use [12]. Previous research shows that diffusion models can eliminate embedded information while preserving utility, supporting diffusion-based sanitization [13], [14].

Sanitization, also known as active steganalysis, involves extracting hidden data from steganographic images while preserving the cover image's quality. Traditional approaches focus on removing this hidden information by tweaking pixel values or modifying the image's frequency components, similar to classic steganography methods. A key challenge lies in the defender's frequent lack of knowledge about the attacker's embedding technique, decoding method, or specific model. Thus, the sanitizer must work quickly and effectively despite this uncertainty.

We introduce Adversarial Diffusion Sanitization (ADS) to address this need. In the scenario shown in Fig. 1, an attacker creates an image carrier using an unknown diffusion steganography method and key; the platform then applies ADS to the images before storing or sharing them; at the receiver end, either decoding fails or results in high bit errors. ADS is designed to be brief: it performs a single forward noising step, adds a targeted adversarial perturbation to the noised data, and then executes one reverse step to generate a visually convincing output. It does not require retraining diffusion models; instead, it utilizes an off-the-shelf diffusion denoiser as a differentiable proxy for various diffusion-based decoders that depend on consistent reverse diffusion or inversion processes. This approach is motivated by evidence that small, carefully optimized perturbations can significantly impair diffusion-model performance in subsequent tasks, even when these perturbations are visually small [15].

A key deployment requirement is preserving utility, including color fidelity and avoiding artifacts that degrade the user experience or interfere with benign vision pipelines. Independent per-channel updates can introduce chromatic artifacts. To mitigate this, we introduce a color-aware update rule that couples RGB channels at each pixel via a quaternion-based direction, improving color consistency without retraining any diffusion model or requiring quaternion-valued diffusion training [16]–[18]. Beyond color stability, ADS is engineered for gateway constraints: bounded iteration counts for predictable latency and throughput; explicit distortion budgets (e.g., SSIM/PSNR/LPIPS) to protect downstream analytics; and content-agnostic operation that prioritizes decoder stability over specific image styles or prompts.

We assess ADS in a realistic threat environment, comparing it with robust diffusion-steganography baselines, including error-correction methods and the practical coverless diffusion approach Pulsar. In this context, payload recovery depends on stable inversion or reverse-diffusion pathways, while the defender employs constrained sanitization without reliable detection. Our findings indicate that at low perceptual distortion, the decoder nearly fails, demonstrating a superior security–utility balance compared with common gateway transformations, such as naive resizing, recompression, or color-space toggling. Although traditional transforms can sometimes disrupt decoding, they often either fall short against modern diffusion stego or cause unacceptable visual quality loss.

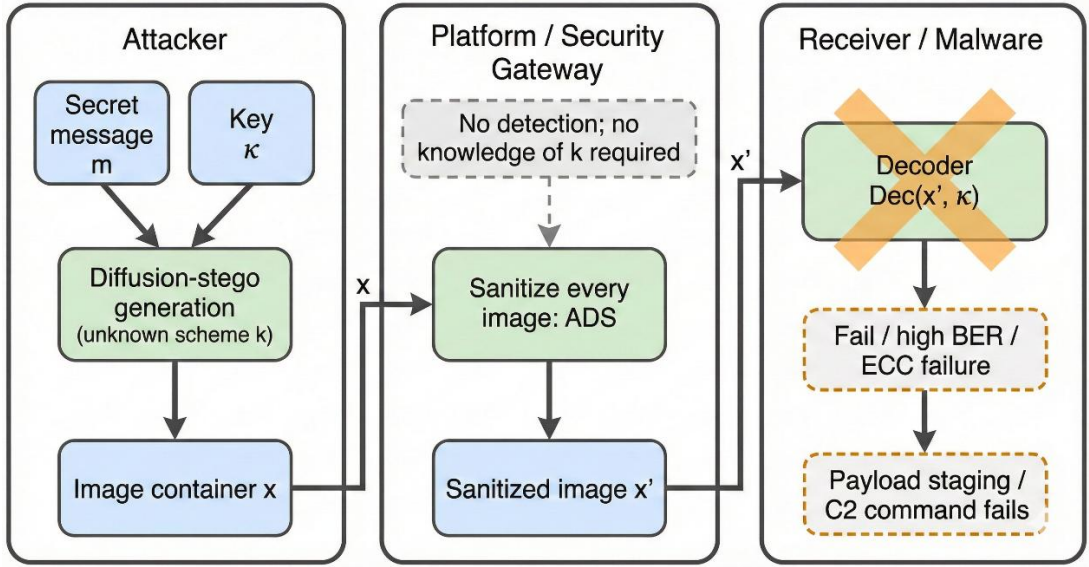

**Figure 1.** Threat model: diffusion-stego as a stegomalware carrier and defense with ADS.

This work makes the following contributions:

1. We develop a Gateway-focused, sanitization-first defense with a strategy that neutralizes payloads without precise detection, suitable for operational scenarios where detection is unreliable or expensive.
2. We introduce ADS, a low-latency, training-free pipeline that performs a targeted update in diffusion space during a single forward–reverse step to disrupt diffusion-based recovery. It uses a standard denoiser as a differentiable proxy for diffusion decoders, simplifying deployment.
3. We propose a quaternion-coupled per-pixel update rule that minimizes color artifacts within small distortion budgets without requiring retraining of any diffusion component. Additionally, it reduces perceptual artifacts while maintaining strict distortion limits and preserving the balance between security and utility.
4. We establish a realistic threat model and compare our approach to robust diffusion-steganography baselines, such as Pulsar. Our results show decoder failures near saturation at low perceptual distortion levels and better security–utility balance than standard gateway transforms.

The remainder of the paper is organized as follows: Section 2 reviews the background and related work. Section 3 describes ADS and the color-aware update rule. ADS addresses (i) the operational gap between detection and neutralization in modern content pipelines; (ii) a model-agnostic deployment approach that uses widely available denoisers and complements traditional steganalysis when reliable detection is available; (iii) the production of sanitized suspect content when detection is infeasible; and (iv) proactive mitigation against diffusion-driven steganographic threats. Section 4 presents the experimental evaluation and deployment considerations. Section 5 concludes the paper.

# 2. Background and Related Work

## 2.1 Steganography and Detection

Networked systems routinely exchange privacy-sensitive content, motivating techniques that conceal not only message content but, in some cases, the existence of communication itself. Information hiding and steganography address this need by embedding secrets into benign-looking carriers so that transmission is less likely to raise suspicion under monitoring [2], [3], [5], [19]–[22]. Steganographic carriers span multiple modalities, including linguistic/text generation [23], images [2], [3], [19], network traffic [20], audio [21], wireless/physical-layer signals [22], and video [24]. Among these, image steganography remains a dominant carrier because images are ubiquitous in web and platform workflows; in the standard cover-modification paradigm, a secret message is embedded into a cover image $X_C$ to form a stego image $X_S$, which is then transmitted between nodes while aiming to remain indistinguishable from benign content to both human observers and statistical detectors [2], [3], [5], [19].

In the classical setting, the main design objectives are: (i) imperceptibility (minimal visual distortion), (ii) statistical security (low detectability by steganalysis), (iii) payload capacity (bits conveyed per pixel or per carrier), and (iv) robustness to processing such as compression, resizing, and format conversion [5], [25]. These objectives align closely with practical adversarial dynamics. An attacker typically pushes toward higher payloads and reliable decoding while attempting to preserve realism and evade detectors, whereas a defender seeks either to detect suspicious carriers or to prevent successful extraction with minimal degradation to benign content utility [6], [25]–[27]. In deployed environments, the defender may also operate as an "active warden" that applies routine transformations (e.g., re-encoding or normalization) to disrupt covert channels even when detection is imperfect or costly to maintain at scale [3], [28].

Traditional image steganography primarily modifies a cover image in the spatial or transform domain. Spatial-domain least-significant-bit (LSB) methods and their variants remain common due to their simplicity and favorable capacity, and have been systematized in prior taxonomies [29]. Capacity-oriented designs extend these ideas using alternative numerical representations and multi bit-plane decompositions to increase payload under controlled distortion constraints [30].

Palette-based schemes introduce different embedding constraints and capacity measures, and pixel sorting has been used to manage embedding choices in indexed-color images [31]. Because such modifications often leave statistical traces, steganalysis has long focused on detecting embedding artifacts using engineered features and learned classifiers. Early examples include modified pixel comparison and complexity-based measures designed to expose LSB-like anomalies [32], while later work introduced substantially stronger feature sets and "rich models" that improved detection performance across many classical schemes [6]. More recently, deep learning has become a standard approach for steganalysis; contemporary surveys summarize the dominant CNN-based families and emphasize the practical gaps that arise under realistic conditions (unknown embedding schemes, unknown payloads, and common post-processing) [26].

Deep learning has also expanded the space of hiding methods. A recent survey of deep-learning-based image steganography categorizes both cover-edited and coverless approaches and highlights a broader shift toward reversible, learnable, and generative designs [33]. This shift is particularly relevant operationally because cryptography and steganography mitigate different aspects of risk: encryption protects message content, but it does not hide the existence of communication and may remain subject to policy enforcement, traffic analysis, or scrutiny in heavily monitored settings [3], [5].

Generative (coverless) steganography instead conceals secrets by synthesizing the carrier image directly from the secret data (often conditioned on a shared key), eliminating the need to modify a specific cover image [9], [34]–[38]. Since no existing image is altered, coverless approaches can reduce the cover, stego discrepancies that many detectors are trained to exploit, and the defender typically lacks a paired "clean" cover for differential analysis, changing assumptions behind a large class of classical steganalysis pipelines [6], [25], [26].

Generative steganography has also become part of a broader security threat model. Steganography can enable stegomalware, where payloads, commands, or keys are hidden in images and moved through normal content channels to support staging and covert command-and-control [4], [28]. This motivates defenses that do not assume static hiding schemes and that remain effective under rapid attacker adaptation, including defenses deployed at content gateways that prioritize disruption and risk reduction over perfect attribution [28].

Historically, GAN-based approaches were among the most widely explored generative steganography methods: secret information can be encoded into conditional inputs, label embeddings, or latent variables and then used to synthesize a carrier image for transmission [9], [34], [35], [39]. However, GAN outputs can exhibit artifacts due to training instability or imperfect distribution matching, which can degrade realism and increase the risk of detection under careful analysis [9], [39]. To address these limitations, subsequent work leveraged alternative generative families that support better control and/or more reliable recovery. Variational autoencoders (VAEs) offer structured latent spaces and principled rate–distortion trade-offs that can simplify encoding [40].

Invertible and flow-based models provide bijective mappings that are attractive when exact or near-exact recovery is required; this includes general normalizing-flow constructions [41], [42] and steganography-specific flow designs that integrate encoding and decoding into an invertible transformation [36], [37]. Flow-style modeling has also continued to evolve through related continuous-time formulations, such as Poisson flow generative models [43]. Autoregressive generators (e.g., PixelCNN/PixelCNN++) provide precise pixel-level control but tend to be computationally expensive at high resolution [44], [45]. Transformer-based vision models support global context modeling via attention and, when paired with discrete image representations, have enabled high-resolution image synthesis pipelines that are useful when hiding must respect semantic or structural constraints [46].

Diffusion models currently dominate high-fidelity image generation and provide additional degrees of control through iterative sampling trajectories and inversion-style formulations [47]–[50]. Advances in sampling efficiency (including exponential-integrator formulations) further increase the practicality of diffusion-based systems in real deployments [51]. These properties have enabled multiple training-free or lightly constrained diffusion steganography schemes that embed information into the initial noise, sampling randomness, or controllable denoising trajectories, while maintaining strong perceptual realism [7], [10], [11].

A common abstraction for generative steganography is that a sender and receiver communicate a secret message $M$ through a generated stego image $X_S$. Unlike classical cover-based steganography [5], the sender uses a generator $G$ to produce the carrier directly from the message (often conditioned on a shared key $K$), and the receiver extracts an estimate $M'$ using an extractor $E$:

$$X_S = G(M, K), M' = E(X_S, K).$$

From a security and deployment viewpoint, this setting changes both attack and defense. The attacker's goal is to produce carriers that remain consistent with natural image distributions while still allowing reliable extraction at the receiver. The defender's goal is to detect stego carriers and/or disrupt extraction while preserving image utility for benign users. In operational environments where detection reliability is limited and adversary behavior evolves rapidly, a practical alternative is to apply bounded-distortion transformations (sanitization) to inbound images to reduce the probability of successful payload recovery at scale [28].

Despite its utility, generative steganography faces persistent technical constraints that also shape defenses:

1. Capacity vs. realism trade-off—high embedding rates can distort the generative process or induce artifacts that increase detectability.
2. Robustness to post-processing—stego images must remain decodable after compression, resizing, filtering, and platform-specific transformations.
3. Recovery reliability and invertibility—extraction must remain stable under noise and transmission loss; many systems leverage invertible mappings, error-control coding, or inversion-capable generative processes to stabilize decoding [42], [47], [49].
4. Statistical indistinguishability—the generated carrier should remain consistent with natural image statistics to reduce exposure to steganalysis and distribution tests [25].

These constraints also delimit defensive strategy: classical steganalysis is strongest when stable cover–stego differences exist and can be learned, whereas coverless generation and distribution-preserving mappings can reduce the strength of that training signal, especially when deployment conditions diverge from training assumptions.

Defense research is commonly grouped into detection-first and sanitization-first strategies. Detection-first defenses attempt to flag suspicious carriers using steganalysis models, including feature-rich detectors and deep detectors trained on known hiding algorithms and payload rates. These approaches remain important but face practical gaps when hiding schemes change, when carriers are synthesized rather than edited, or when post-processing shifts the statistics of both benign and malicious content. Sanitization-first defenses instead treat inbound images as untrusted and apply bounded-distortion transformations intended to break extraction while preserving utility (e.g., re-encoding, resizing, and filtering). Recent research has started to examine diffusion-driven sanitization methods that use pretrained diffusion components as differentiable proxies to interfere with diffusion-based hiding and decoding. This approach is critical when the threat model involves training-free diffusion steganography and stegomalware delivery.

Diffusion models strengthen coverless (generative) steganography because the carrier image is synthesized rather than edited [4]. A diffusion model generates an image by learning to reverse a gradual noising process. In Denoising Diffusion Probabilistic Models (DDPMs) [47], a forward Markov chain perturbs a clean sample $x_0$ into a noisy sample $x_t$ by adding Gaussian noise using a variance schedule $\{\bar{\alpha}_t\}_{t=1}^T$. This forward process can be written as

$$x_t = \sqrt{\bar{\alpha}_t}\, x_0 + \sqrt{1 - \bar{\alpha}_t}\, \varepsilon, \qquad \varepsilon \sim \mathcal{N}(0, I),$$

where $\bar{\alpha}_t = \prod_{s=1}^t \alpha_s$ denotes the cumulative product of per-step noise-retention coefficients. A neural denoiser, often based on a U-Net, is trained to predict the injected noise $\varepsilon$, or equivalently the score $\nabla_{x_t} \log p(x_t)$. This enables reverse-time updates that move a noisy state back toward the data distribution [47]. Many practical systems use deterministic or near-deterministic samplers that support fast generation and also support inversion, meaning an observed image can be mapped back to a latent or noise state under a chosen scheduler. DDIM [48] provides a deterministic trajectory that supports more stable inversion than stochastic DDPM sampling. Score-based formulations connect diffusion models to stochastic differential equations (SDEs) and probability-flow ordinary differential equations (ODEs), which provide explicitly invertible dynamics [49]. Latent Diffusion Models (LDMs) move the diffusion process to a compressed latent space while keeping high visual fidelity [50], which matters because many recent hiding and defense methods are built around Stable-Diffusion-style pipelines.

Diffusion-based coverless steganography uses the diffusion process as the carrier for secret information. The main difference across methods is where the secret is placed in the pipeline and what the decoder expects during recovery. Some schemes encode bits in the initial noise (or latent) seed before sampling. Diffusion-Stego [7] is a representative example that projects the message into the noise or latent space in a training-free way and then samples an image, allowing trade-offs between image quality and extraction accuracy. Other schemes encode bits across sampling steps by controlling the stochasticity used during generation. Pulsar [52] follows this approach by selecting variance-noise streams using a key, then applying error-correcting codes (ECC) to tolerate partial corruption. This is important for defense because ECC can convert moderate perturbations into correct recovery. A third family builds an explicitly reversible mapping between secrets and diffusion latent and then relies on an invertible sampling path. Plug-and-Hide [53] formalizes this direction by combining a bit-to-Gaussian mapping with an invertible probability-flow ODE, enabling adjustable trade-offs among image quality, extraction accuracy, and security. Related constructions also use invertible diffusion and ODE mappings to support reversible secret–image conversion [54].

Conditional diffusion introduces additional degrees of freedom that can carry data, including prompts and auxiliary conditioning. CRoSS [10] uses conditional diffusion translation and control mechanisms to provide more controllable and robust hiding. DiffStega [11] addresses weaknesses of prompt-only keys by using a password-dependent reference image as an image prompt and adding a "Noise Flip" mechanism to increase password sensitivity. LDStega extends this direction for latent diffusion by introducing latent-space coding strategies intended to improve controllability and robustness under common transforms such as JPEG [55]. Table 1 summarizes representative diffusion-based schemes and their primary embedding locations.

These developments make detection-first defenses less reliable in practice. When the carrier is generated rather than modified, many classical detection assumptions based on cover–stego differences are harder to apply directly [2], [3]. Several diffusion-based designs also aim to keep latent or noise variables close to their expected priors, which reduces the power of distributional tests [53]. In addition, schemes can change their embedding location and recovery path across designs, which increases the cost of maintaining fixed detectors in a large deployment [7], [10], [11], [52], [53]. This motivates gateway defenses that do not require accurate scheme identification before applying protection.

**Table 1.** Representative diffusion-based steganography schemes.

| Scheme | Carrier type | Primary embedding locus | Key idea (high level) |
|---|---|---|---|
| CRoSS [10] | Text-to-image container | Prompt/key conditioning; diffusion translation/control | Controllable/robust hiding using conditional diffusion priors |
| Diffusion-Stego [7] | Generated image | Initial noise (message projection) | Training-free message projection into noise/latent space; adjustable security–utility trade-offs |
| DiffStega [11] | Coverless (password + reference image) | Image prompt + "Noise Flip" | Password-dependent image prompt; noise flip to harden unauthorized recovery |
| Pulsar [52] | Generated image | Variance-noise streams + ECC | Bit encoding via keyed variance-noise selection; ECC decoding for robustness |
| Plug-and-Hide [53] | Generated image | Bit-to-Gaussian mapping + invertible ODE | Provable reversible mapping; probability-flow ODE inversion |
| LDStega [55] | Generated image (latent diffusion) | Latent-space coding (LDM) | Latent-domain embedding with controllability and robustness to common transforms |
| GSD / StegoDiffusion [54] | Generated image | Invertible diffusion + ODE mapping | Invertible diffusion construction enabling reversible secret–image conversion |

Sanitization defenses instead transform inputs to reduce the ability to recover hidden content while preserving user-perceived utility. Prior work includes heuristic transforms such as re-encoding, resizing, denoising, and filtering, as well as learned sanitizers trained to remove hidden signals. Earlier studies explored machine-learning sanitization for images that contain stegomalware [56], and more recent work continues to develop deep-learning sanitizers designed to disrupt embedded payloads under visual constraints [28]. Open Image Content Disarm and Reconstruction (ICDR) proposes a practical CDR-style pipeline for JPEG images that removes non-image structures and metadata and manipulates pixel data to disable embedded threats while keeping usability [57].

Recent work also shows that diffusion models can be used as sanitizers. SUDS [32] proposes a learned framework that targets multiple hiding regimes without requiring explicit knowledge of the hiding technique. DM-SUDS extends diffusion-based sanitization to higher-complexity data and studies the trade-off between "safety" - secret removal and "utility" - fidelity of the image [13]. Related work shows that some hidden signals, such as invisible watermarks, are removable using generative AI under suitable conditions [14]. Although these methods mainly focus on image-into-image hiding, they support diffusion-based processing as a viable approach for removing embedded information under perceptual constraints.

A closely related research direction studies adversarial perturbations that disrupt diffusion-model behavior. PhotoGuard introduces small perturbations that cause diffusion-based image editing to fail or produce implausible outputs, effectively protecting images against malicious edits [15]. AdvDM similarly uses adversarial examples to prevent unwanted mimicry such as style imitation [58]. DiffusionGuard argues that targeting early denoising stages can improve effectiveness and robustness, including under masked editing, and proposes objectives and augmentations to improve transferability [59]. This literature indicates that carefully optimized, low-magnitude perturbations can cause large downstream effects in diffusion pipelines. It also suggests that practical deployments may benefit from randomization, model rotation, and complementary transforms in an arms-race setting [60].

Finally, any security sanitization pipeline must preserve utility, including color fidelity. If RGB channels are perturbed independently, small distortions can appear as chromatic speckle or channel-wise artifacts. Quaternion representations model an RGB pixel as a coupled entity and therefore motivate channel-coupled operations [61]–[63]. Quaternion neural networks have been shown to capture cross-channel correlations with compact parameterizations in several settings [16], [17], [64], and quaternion-based methods have been applied to color image classification and forensics [18]. These properties are relevant for security gateways where latency and predictable computation are essential constraints.

## 2.2 Quaternions for Color Image Processing

Recent advances, such as the quaternion gradient, quaternion Fourier transform, HR-calculus, and GHR calculus, have led to increasing use of quaternions and other hyper-complex algebras in signal processing and artificial intelligence. Proper feature representation of input data is essential for successful machine learning. For example, quaternion algebra naturally expresses three-dimensional rotations, avoiding the problem of gimbal lock that occurs with standard vector arithmetic. The color of an image pixel, as a triplet of intensities of (R)ed, (G)reen, and (B)lue channels, can be naturally represented by quaternions. This makes quaternion representation particularly useful in computer vision and image processing.

A quaternion number $q \in \mathbb{H}$ extends the concept of complex numbers by introducing one real (a) and three imaginary (b, c, d) components in the form: $q = a + bi + cj + dk$; here $a, b, c, d \in \mathbb{R}$ and $(i, j, k)$ form the quaternion unit basis, where $i^2 = j^2 = k^2 = ijk = -1$. An algebra on $\mathbb{H}$ defines operations on quaternion numbers, such as addition, conjugation, and norm, similarly to the algebra on complex numbers [63]. For a quaternion $q = a + bi + cj + dk$, operations are defined as follows:

Multiplication by a scalar $\lambda \in \mathbb{R}$: $\lambda q = \lambda a + \lambda bi + \lambda cj + \lambda dk$
Conjugation: $q^* = a - bi - cj - dk$
Norm: $|q| = \sqrt{q \otimes q^*} = \sqrt{a^2 + b^2 + c^2 + d^2}$
For two quaternions $q_1 = a_1 + b_1 i + c_1 j + d_1 k$ and $q_2 = a_2 + b_2 i + c_2 j + d_2 k$:
Addition: $q_1 + q_2 = (a_1 + a_2) + (b_1 + b_2)i + (c_1 + c_2)j + (d_1 + d_2)k$
Multiplication (Hamilton product):
$$q_1 \otimes q_2 = (a_1 a_2 - b_1 b_2 - c_1 c_2 - d_1 d_2)$$

$$+ (a_1b_2 + b_1a_2 + c_1d_2 - d_1c_2)i$$
$$+ (a_1c_2 - b_1d_2 + c_1a_2 + d_1b_2)j$$
$$+ (a_1d_2 + b_1c_2 - c_1b_2 + d_1a_2)k$$

Hamilton products are noncommutative, e.g.: $q_1 \otimes q_2 \neq q_2 \otimes q_1$, $q_1, q_2 \in \mathbb{H}$. Hamilton introduced quaternions in 1843.

Other algebras were introduced later, [65]–[68]. There are 4 possible algebras with three imaginary components that differ in the relations between components and as a result definition of the Hamilton product (Table 2): bi-quaternions, double-complex, and HCA4. Most of the works on hypercomplex neural networks employ quaternions, but in most cases, alternative algebras could be used as well. There is limited literature on the use of other algebras in machine learning and machine learning, such as Cayley-Dickson algebras, Clifford Algebras, MacFarlane's Hyperbolic Quaternions, Klein four-group numbers [69].

**Table 2.** Hypercomplex algebras with 3 imaginary components

| Quaternions | Reduced bi-quaternions | Double-complex | HCA4 |
|---|---|---|---|
| **Relations among imaginary components i, j, k:** | | | |
| $\begin{cases} i^2 = j^2 = k^2 = ijk = -1 \\ ij = -ji = k \\ jk = -kj = i \\ ki = -ik = j \end{cases}$ | $\begin{cases} i^2 = k^2 = -1 \\ j^2 = 1 \\ ij = ji = k \\ jk = kj = i \\ ki = ik = -1 \end{cases}$ | $\begin{cases} j^2 = k^2 = -1 \\ i^2 = 1 \\ ij = ji = k \\ jk = kj = -i \\ ki = ik = j \end{cases}$ | $\begin{cases} i^2 = j^2 = -1 \\ k^2 = 1 \\ ij = ji = -k \\ jk = kj = i \\ ki = ik = j \end{cases}$ |
| **Hypercomplex product $x \otimes y$ in matrix form, $q_1 = a_1 + b_1i + c_1j + d_1k$ and $q_2 = a_2 + b_2i + c_2j + d_3k$.** | | | |
| $\begin{bmatrix} a_1 & -b_1 & -c_1 & -d_1 \\ b_1 & a_1 & d_1 & -c_1 \\ c_1 & -d_1 & a_1 & b_1 \\ d_1 & c_1 & -b_1 & a_1 \end{bmatrix} \begin{bmatrix} a_2 \\ b_2 \\ c_2 \\ d_2 \end{bmatrix}$ | $\begin{bmatrix} a_1 & -b_1 & c_1 & -d_1 \\ b_1 & a_1 & d_1 & c_1 \\ c_1 & -d_1 & a_1 & -b_1 \\ d_1 & c_1 & b_1 & a_1 \end{bmatrix} \begin{bmatrix} a_2 \\ b_2 \\ c_2 \\ d_2 \end{bmatrix}$ | $\begin{bmatrix} a_1 & b_1 & -c_1 & -d_1 \\ b_1 & a_1 & -d_1 & -c_1 \\ c_1 & d_1 & a_1 & -b_1 \\ d_1 & c_1 & b_1 & a_1 \end{bmatrix} \begin{bmatrix} a_2 \\ b_2 \\ c_2 \\ d_2 \end{bmatrix}$ | $\begin{bmatrix} a_1 & -b_1 & -c_1 & d_1 \\ b_1 & a_1 & -d_1 & c_1 \\ c_1 & d_1 & a_1 & -b_1 \\ d_1 & -c_1 & -b_1 & a_1 \end{bmatrix} \begin{bmatrix} a_2 \\ b_2 \\ c_2 \\ d_2 \end{bmatrix}$ |

Quaternions are similar to 3- or 4-dimensional vectors, but they are processed differently, which makes them suitable for algorithms that operate on hypercomplex data. In the following subsections, we elaborate on how quaternions could be used to represent color images, introduce quaternion neurons, and other components of quaternion neural networks.

Typically, color is represented using a color space such as RGB, HSV, or Lab. [63]. For example, the RGB color space represents color as the intensity of three base colors: red, green, and blue. The color of a single pixel is therefore represented as a triplet (r, g, b), where r, g, and b are numbers in the range of [0,1]. The color image of size H by W pixels is represented as a set of three matrices R, G, and B. Alternatively, the color of a single pixel could be represented as a single quaternion in the form: $q = 0 + ri + gj + bk$. In the general case, the color image of the size H by W pixels is represented as a quaternion matrix $I \in \mathbb{H}^{H \times W}$:

$$I = 0 + Ri + Gj + Bk$$

where $R, G, B \in \mathbb{R}^{H \times W}$ are real-valued matrices representing red, green, and blue channels, respectively. This representation opens rich possibilities for color image processing and the design of convolutional neural networks. Not only can the input image be represented in this form, but intermediate features can as well, leading to better preservation of the interrelationships and structural information between the various channels of the feature map.

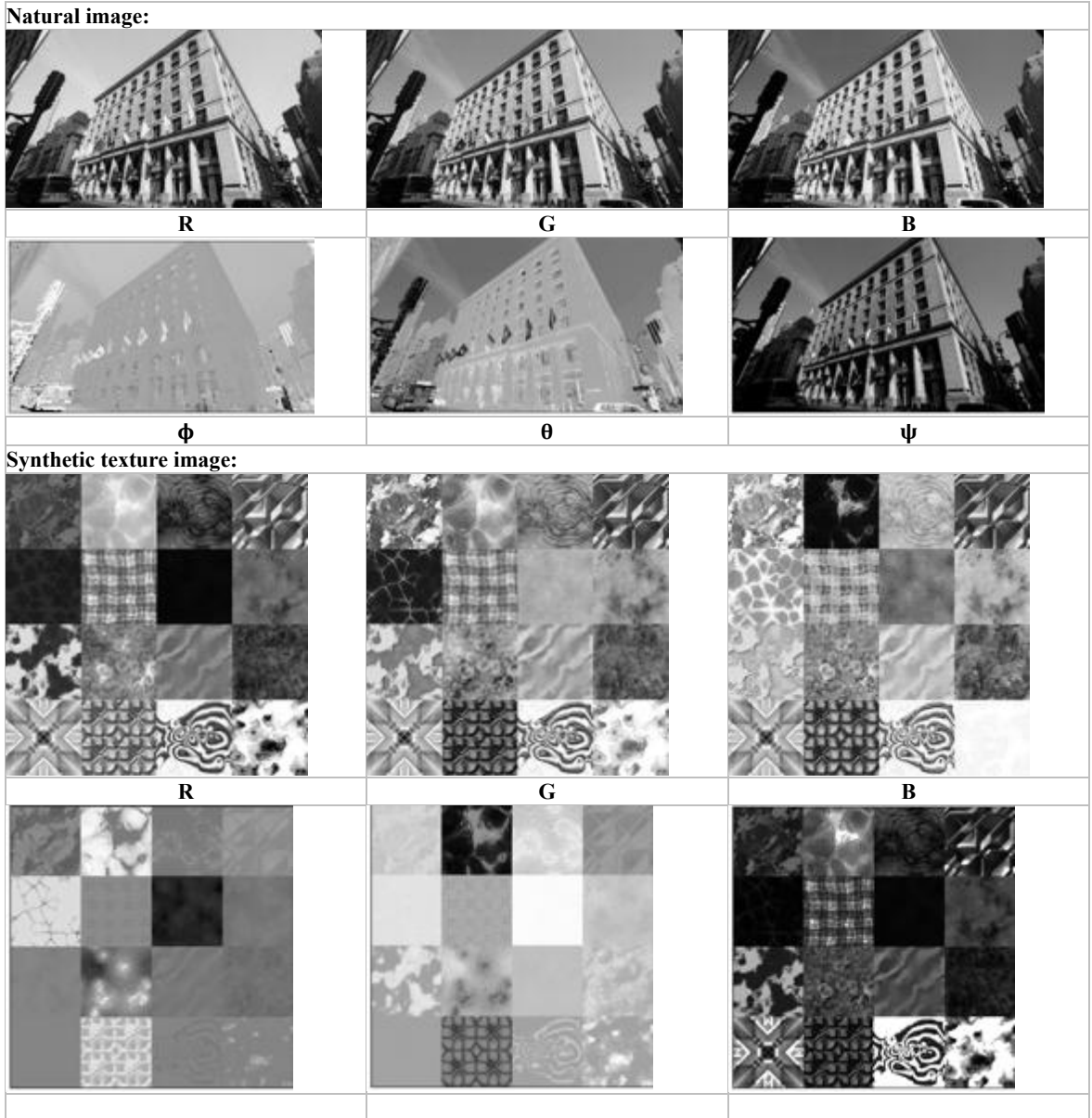

**Figure 2.** Image representation (RGB channels on top and phase components in the bottom)

Contrary to the traditional representation, quaternions could be represented using separate magnitudes and phases. In general, a quaternion $q = a + bi + cj + dk$ could be represented in the form [70]:

$$q = \|q\|e^{i\phi}e^{j\theta}e^{k\psi}$$

where $\|q\|$ is the magnitude; $\phi = \text{atan2}(n_\phi, d_\phi)$, $\theta = \text{atan2}(n_\theta, d_\theta)$, $\psi = \arcsin(n_\psi)$ capture the phase information. Here

$$n_\varphi = 2(cd + ab)$$

$$d_\varphi = a^2 - b^2 + c^2 - d^2$$
$$n_\theta = 2(bd + ac)$$
$$d_\theta = a^2 + b^2 - c^2 - d^2$$
$$n_\psi = 2(bc + ad)$$

This is especially useful for multimodal input data, such as a combination of an RGB image and depth information. For example, at work [71] reduced biquaternions are used to encode color and depth information simultaneously. The difference between RGB and phase representation is presented in Fig. 2., magnitude images are shown in Fig. 3.

Magnitude-image looks similar to the grayscale image, but numerically they are different as root-square is involved in the computation of the magnitude.

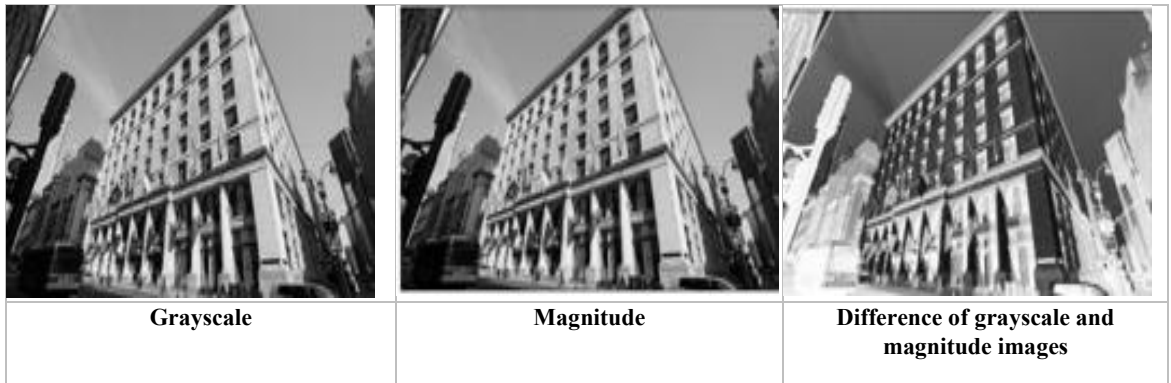

| Grayscale | Magnitude | Difference of grayscale and magnitude images |

**Figure 3.** Grayscale image and magnitude image.

The quaternion multiplication $I_1 \otimes I_2$ combines the inputs in an entirely distinct manner in comparison with real point-wise multiplication $I_1 \cdot I_2$, as it involves the cross product of the imaginary components of the quaternions. As can be seen, operations $I_1 \otimes I_2$ and $I_2 \otimes I_1$ produce different results due to the non-commutativity of the Hamilton product.

This representation could serve as a primary feature extraction step in neural networks. Beyond this, the quaternion representation offers additional tools for image processing. Operations on input images $I_1$ and $I_2$ are illustrated in Fig. 4

Summation operation $I_1 + I_2$ is equivalent for quaternion and real cases. Conjugation $I_1^*$ involves flipping the sign of the three imaginary components, which is equivalent to inverting the sign of the corresponding vector components in the geometric interpretation of quaternions. In image processing, conjugation can be used to perform operations such as image inversion.

| | Real | I | J | K |
|---|---|---|---|---|
| $I_1$ | | | | |
| $I_2$ | | | | |
| $I_1^*$ (quaternion conjugation) | | | | |
| $I_1 + I_2$ | | | | |
| $I_1 \cdot I_2$ (real pointwise-multiplication) | | | | |
| $I_1 \otimes I_2$ (quaternion Hamilton product) | | | | |
| $I_2 \otimes I_1$ (quaternion Hamilton product) | | | | |

**Figure 4.** Quaternion image algebra operations

## 3. The method

We consider a security gateway that sanitizes every incoming image without first deciding whether it contains a hidden message. Let the input be an RGB image $x_0 \in [-1,1]^{H \times W \times 3}$. An attacker may generate $x_0$ using an unknown diffusion-based steganography scheme $k$ and key $\kappa$, such that a receiver-side decoder $\text{Dec}_{k,\kappa}(\cdot)$ can recover an estimated message $\hat{m}$. The gateway applies a sanitization function $f$ and outputs $x' = f(x_0)$.

The goal is to maximize decoding disruption while keeping image distortion bounded. At a high level, we target the following security–utility objective:

$$\max_{f \in \mathcal{F}} \mathbb{E}_{\text{Dec} \sim \mathcal{D}} \left[ \text{Fail}\left(\text{Dec}(f(x_0))\right) \right] \quad \text{s.t.} \quad \text{Dist}(x_0, f(x_0)) \leq \tau,$$

where $\mathcal{D}$ is a threat set of plausible decoders Dec (unknown to the defender), $\text{Fail}(\cdot)$ indicates decoder failure, and $\text{Dist}(\cdot,\cdot)$ is a distortion measure such as PSNR or SSIM.

Since Dec is unknown, we optimize a differentiable surrogate based on a shared dependency across many diffusion-steganography designs: recovery often relies on stable inversion or consistent reverse diffusion. ADS uses an off-the-shelf diffusion denoiser only as a differentiable proxy for this behavior.

## 3.1 Adversarial Diffusion Sanitization

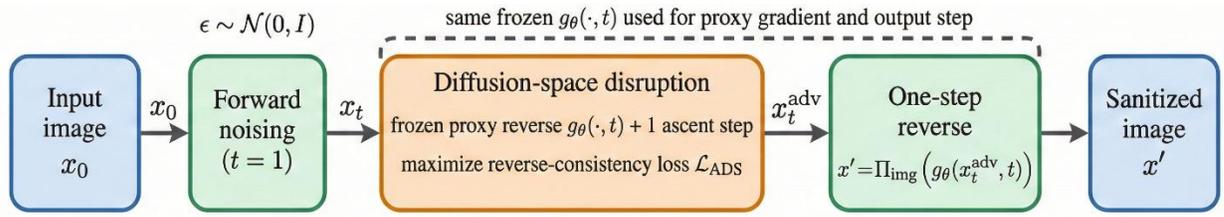

**Figure 5.** ADS mechanism: one-step forward diffusion, diffusion-space adversarial update, and one-step reverse diffusion.

Figure 5 summarizes ADS, details are presented in Algorithm 1. The pipeline is designed to be short to comply with gateway latency limits. It performs one forward noising step, executes a single adversarial update in diffusion space to disrupt reverse consistency, and then takes one reverse step to revert to image space.

In brief, ADS first enters diffusion space by sampling noise and applying a single forward diffusion step to obtain $x_t$ (Steps 1–2). It then applies the fixed proxy denoiser to compute a one-step reverse estimate $g_\theta(x_t, t)$ (Step 3). Next, it maximizes a reverse-consistency loss by taking one diffusion-space ascent update with projection/clamping to control distortion and ensure numerical stability (Steps 4–7). Finally, it returns to image space by applying the same one-step reverse mapping to the adversarial state to produce the sanitized output $x'$ (Step 8). The details of each part are given below.

**Forward diffusion (Steps 1-2):** Let $\bar{\alpha}_t$ denote the cumulative noise schedule, with $\bar{\alpha}_t = \prod_{s=1}^{t} \alpha_s$. For a small diffusion time $t$ near the clean endpoint (typically $t = 1$ in discrete schedules), we sample $\varepsilon \sim \mathcal{N}(0, I)$ and compute

$$x_t = \sqrt{\bar{\alpha}_t}\, x_0 + \sqrt{1-\bar{\alpha}_t}\, \varepsilon.$$

This step moves the image into a nearby noised state where one-step denoising is well-defined, while keeping perceptual distortion small.

| Algorithm 1. Adversarial Diffusion Sanitization (ADS) | |
|---|---|
| **Inputs:** $x_0 \in [-1,1]^{H \times W \times 3}$; proxy denoiser $\epsilon_\theta(\cdot, t)$ with fixed $\theta$; cumulative schedule $\bar{\alpha}_t$; small timestep $t$; step size $\varepsilon_{\text{adv}}$; projection/clamp $\Pi(\cdot)$; (optionally) stability constant $\delta$ for the quaternion-aware variant. | |
| **Output:** sanitized image $x'$. | |
| 1: Sample $\varepsilon \sim \mathcal{N}(0, I)$ | Randomizes the diffusion entry point |
| 2: $x_t \leftarrow \sqrt{\bar{\alpha}_t}\, x_0 + \sqrt{1-\bar{\alpha}_t}\, \varepsilon$ | Single forward noising step. |
| 3: Compute $\hat{x}_0 \leftarrow g_\theta(x_t, t)$ | One-step reverse estimate via proxy. |
| 4: $\mathcal{L}_{\text{ADS}} \leftarrow \dfrac{1}{3HW}\|\hat{x}_0 - x_0\|_2^2$ | Maximize |
| 5: $G \leftarrow \nabla_{x_t} \mathcal{L}_{\text{ADS}}$ | Backprop through proxy; $\theta$ remains frozen. |
| 6: Compute update $\Delta_t$ | FGSM: $\Delta_t = \varepsilon_{\text{adv}} \text{sign}(G)$. QDir (Sec. 3.2): per-pixel coupled direction. |
| 7: $x_t^{\text{adv}} \leftarrow \Pi(x_t + \Delta_t)$ | Projection/clamp for numerical stability and bounded distortion. |
| 8: $x' \leftarrow \Pi_{\text{img}}\!\left(g_\theta(x_t^{\text{adv}}, t)\right)$ | Single reverse step |

**One-step reverse mapping as a proxy (Step 3):** Let $\epsilon_\theta(x_t, t)$ be the proxy denoiser prediction of the injected noise at time $t$, with fixed parameters $\theta$. A standard one-step estimate of the clean image implied by the proxy is

$$g_\theta(x_t, t) := \hat{x}_0(x_t, t) = \frac{x_t - \sqrt{1-\bar{\alpha}_t}\, \epsilon_\theta(x_t, t)}{\sqrt{\bar{\alpha}_t}}.$$

In practice, $g_\theta$ can include the denoiser together with the chosen scheduler's one-step update (DDIM- or DDPM-style). In the small-$t$ regime, $g_\theta(\cdot, t)$ serves as a fast, differentiable proxy for reverse-consistency.

**Disruption loss and diffusion-space adversarial update (Steps 4-7):** ADS aims to make the proxy reconstruction inconsistent with the original input. We therefore maximize the reconstruction-consistency loss

$$\mathcal{L}_{\text{ADS}}(x_t; x_0) = \frac{1}{3HW} |g_\theta(x_t, t) - x_0|_2^2.$$

Let $G = \nabla_{x_t} \mathcal{L}_{\text{ADS}}(x_t; x_0)$. ADS performs a single ascent step in diffusion space to obtain an adversarial state $x_t^{\text{adv}}$. The simplest variant uses an $\ell_\infty$-style FGSM update:

$$x_t^{\text{adv}} = \Pi\!\left(x_t + \varepsilon_{\text{adv}} \text{sign}(G)\right),$$

where $\varepsilon_{\text{adv}} > 0$ controls the security–utility trade-off and $\Pi(\cdot)$ is a projection/clamping operator that keeps values in the numerical range expected by the proxy implementation.

**Single-step return to image space (Step 8):** Finally, ADS produces the sanitized output using the same one-step reverse mapping:

$$x' = f(x_0) = \Pi_{\text{img}}\left(g_\theta(x_t^{\text{adv}}, t)\right),$$

where $\Pi_{\text{img}}$ clamps to the valid image range (e.g., $[-1,1]$ prior to de-normalization).

This forward-noise → adversarial update → one-step reverse structure keeps compute bounded (two denoiser forward passes and one backward pass) while targeting a mechanism that many diffusion-stego decoders depend on: stable inversion and trajectory consistency.

### 3.2 Quaternion-aware coupled update

A practical gateway constraint is color fidelity. Under tight distortion budgets, per-channel sign updates can introduce chromatic speckle because the $R$, $G$, and $B$ channels are perturbed independently. ADS-QDir addresses this by coupling the RGB channels at each pixel during the adversarial update, without retraining the proxy denoiser and without using quaternion-valued diffusion models.

A quaternion is represented as [72] :

$$q = a + b\mathbf{i} + c\mathbf{j} + d\mathbf{k}, \qquad \mathbf{i}^2 = \mathbf{j}^2 = \mathbf{k}^2 = \mathbf{ijk} = -1.$$

An RGB pixel $(r, g, b)$ is treated as a pure quaternion

$$q_{\text{rgb}} = 0 + r\mathbf{i} + g\mathbf{j} + b\mathbf{k}.$$

Let $G(u, v) = (G_r, G_g, G_b)$ be the gradient of $\mathcal{L}_{\text{ADS}}$ with respect to the three channels at pixel $(u, v)$. The corresponding pure-quaternion gradient is

$$G_q(u, v) = 0 + G_r\mathbf{i} + G_g\mathbf{j} + G_b\mathbf{k}.$$

Instead of applying $\text{sign}(G)$ element-wise, ADS-QDir uses a normalized coupled direction:

$$\text{QDir}(G_q) = \frac{G_q}{|G_q|_2 + \delta}, \qquad |G_q|_2 = \sqrt{G_r^2 + G_g^2 + G_b^2},$$

where $\delta > 0$ is a small constant for numerical stability. In RGB form, the per-pixel update becomes

$$\Delta_t(u, v) = \varepsilon_{\text{adv}} \frac{G(u, v)}{|G(u, v)|_2 + \delta},$$

and the diffusion-state update is

$$x_t^{\text{q-adv}}(u, v) = \Pi\left(x_t(u, v) + \Delta_t(u, v)\right).$$

The sanitized output is computed as:

$$x' = \Pi_{\text{img}}\left(g_\theta(x_t^{\text{q-adv}}, t)\right).$$

Operationally, ADS-QDir changes only Step 6 in Algorithm 1. The update direction is shared across channels at each pixel (scaled by the local RGB gradient direction), which reduces channel-wise discontinuities while keeping the same objective of disrupting reverse-diffusion consistency.

# 4. Experiments

We evaluate ADS in an end-to-end diffusion-steganography setting using Pulsar Pulsar is a particularly relevant target because it combines diffusion sampling with error-correction code recovery, which is designed to tolerate moderate corruption and therefore serves as a strong baseline for gateway sanitization. All results in this section are obtained without training, using only pretrained diffusion components. All experiments were run on an RTX 5080 GPU, Ryzen 9 9950X3D CPU, and 64 GB RAM computer using Python 3.10 with PyTorch 2.9 and diffusers 0.36, and the official Pulsar implementation.

## 4.1 Experimental setup

We generate 64 images carrying a hidden message with a resolution of $256 \times 256$ using Pulsar. Publicly available backbone "google/ddpm-church-256" and a DDIM sampler with 50 inference steps are used [48]. Each image encodes an 8192 byte payload under a fixed key $\kappa$ consisting of 64 symbols "E". Sanitization is applied to these carrier images, and the Pulsar decoder is then run on the sanitized outputs. To avoid conflating sanitization with file-format re-encoding, we pass tensors directly into the decoder path; sanitized images are additionally saved as 16-bit PNG purely for inspection.

We evaluate the following sanitization baselines: identity, JPEG recompression ($Q \in \{90,70\}$), resize ($256 \rightarrow 224 \rightarrow 256$), Gaussian blur ($\sigma \in \{0.5,1.0\}$), a diffusion "1-step" baseline (one forward noising step followed by one reverse step, with no adversarial update), and ADS at a single diffusion timestep $t = 1$. ADS uses the same pretrained backbone as a differentiable proxy. It performs one forward diffusion step, applies either a standard FGSM-style update (ADS-FGSM) or the quaternion-coupled per-pixel direction update (ADS-QDir), and then returns to image space with one reverse step, as described in Section 3. Unless otherwise stated, we clamp all intermediate states and outputs to $[-1,1]$. For ADS, we sweep among $\epsilon_{adv} \in \{0,0.005,0.01,0.02\}$; for ADS-QDir we set $\delta = 10^{-8}$. We estimate security using the Pulsar decoding success rate (DSR), the failure rate $(100 \cdot (1-DSR))$, the bit error rate (BER) when a bitstring is produced, and the ECC failure rate reported by Pulsar. Since long payloads can make strict "perfect decode" criteria saturate in small-N studies, we count a decode as successful when $BER \leq 0.48$. Empirically, this threshold matches ECC success in this run while preserving a non-trivial operating range across baselines. Utility is measured by PSNR, SSIM [73], and LPIPS [74] between the sanitized output and the input carrier image.

## 4.2 Quantitative results

Table 3 reports the primary comparison for a low-distortion ADS configuration ($t = 1, \epsilon_{adv} = 0.01$), i.e., an early diffusion step with a small adversarial step size. Two observations follow. First, entering diffusion space alone is not sufficient: the diffusion 1-step baseline preserves the carrier (LPIPS $= 8 \times 10^{-4}$, SSIM $= 0.9979$) and maintains high decoder success (DSR $= 0.9219$). This indicates that a near-identity forward–reverse pass at $t = 1$ does not significantly affect the information Pulsar uses for decoding.

Second, adding the adversarial update in diffusion space produces a large security gain at low distortion. With $\epsilon_{adv} = 0.01$, ADS-FGSM reduces DSR to 0 (100% failure) at LPIPS $= 0.0048$. ADS-QDir achieves 98.4% failure at lower distortion (LPIPS $= 0.0022$) and with higher PSNR/SSIM, giving a better security–utility trade-off at essentially the same security level. ADS achieves comparable or stronger disruption with lower perceptual distortion. JPEG at $Q = 70$ also yields 100% failure, but requires LPIPS $= 0.0126$ and reduces SSIM to 0.9678. Gaussian blur at $\sigma = 1.0$ reaches 100% failure only at much higher distortion (LPIPS $= 0.2606$). At similar distortion to ADS-QDir (LPIPS $\approx 0.002$–$0.003$), JPEG at $Q = 90$ achieves 89.1% failure, whereas ADS-QDir achieves 98.4% failure while also improving PSNR and SSIM.

**Table 3.** End-to-end Pulsar decoding after sanitization ($N = 64$, $256 \times 256$). ADS uses $t = 1$, $\delta = 10^{-8}$ for QDir.

| Method | Parameters | Failure rate (%) | BER | PSNR | SSIM | LPIPS | Latency (ms) |
|---|---|---|---|---|---|---|---|
| Identity | – | 1.56 | 0.2881 | $\infty$ | 1.000 | 0.0000 | 0.31 |
| JPEG | $Q = 90$ | 89.06 | 0.4852 | 43.43 | 0.989 | 0.0028 | 1.84 |
| JPEG | $Q = 70$ | 100.00 | 0.4929 | 37.47 | 0.967 | 0.0126 | 1.65 |
| Resize | $256 \to 224 \to 256$ | 59.37 | 0.4817 | 36.26 | 0.964 | 0.0647 | 2.01 |
| Blur | $\sigma = 0.5$ | 12.50 | 0.4713 | 38.36 | 0.980 | 0.0329 | 2.13 |
| Blur | $\sigma = 1.0$ | 100.00 | 0.4902 | 30.44 | 0.870 | 0.2606 | 2.13 |
| Diffusion 1-step | $t = 1, \epsilon_{adv} = 0$ | 7.81 | 0.4698 | 50.54 | 0.997 | 0.0008 | 27.72 |
| ADS-FGSM | $t = 1, \epsilon_{adv} = 0.01$ | 100.00 | 0.4946 | 43.80 | 0.988 | 0.0048 | 82.75 |
| ADS-QDir | $t = 1, \epsilon_{adv} = 0.01$ | 98.4375 | 0.4857 | 46.31 | 0.994 | 0.0022 | 82.86 |

## 4.3 Qualitative comparison

Fig. 6 shows representative qualitative results at $\epsilon_{adv} = 0.01$. Across the examples, ADS outputs remain visually plausible and close to the input carriers, consistent with the SSIM values in Table 2. At $\epsilon_{adv} = 0.01$, both ADS variants achieve similar decoder disruption, but ADS-QDir yields lower LPIPS and higher SSIM than ADS-FGSM. The qualitative panels also show fewer channel-wise "grain" artifacts in smooth-color areas, consistent with the per-pixel RGB channel coupling in the QDir update.

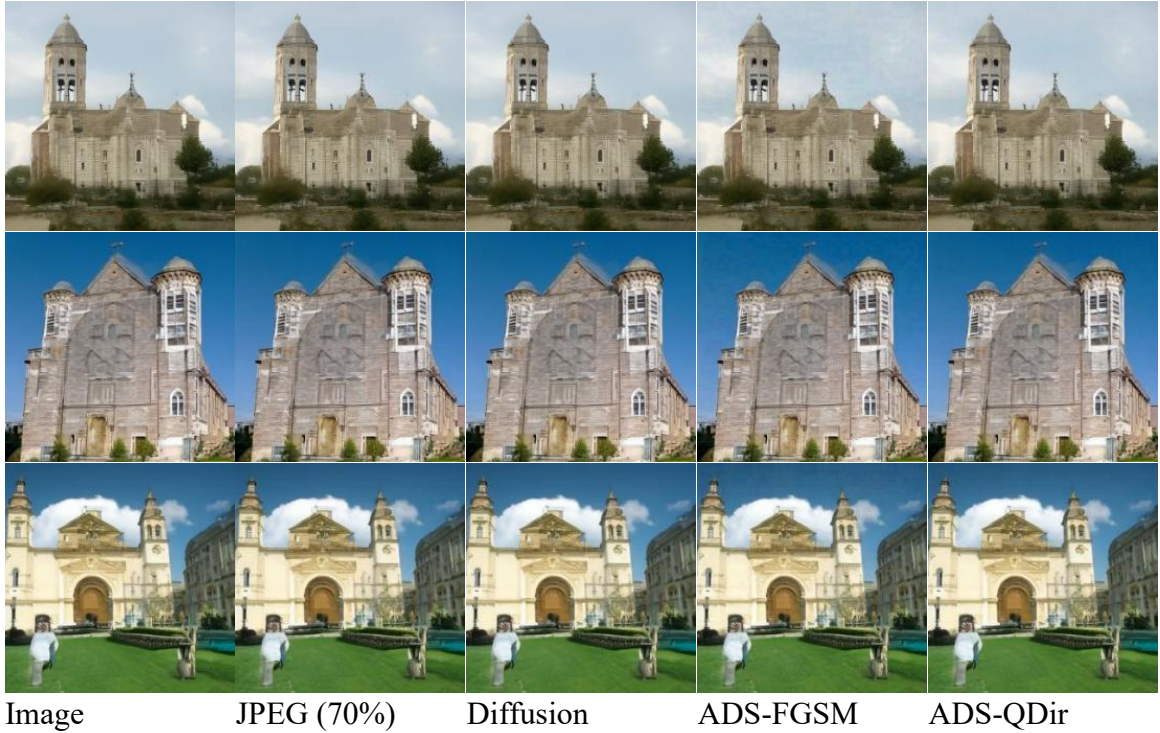

| Image | JPEG (70%) | Diffusion | ADS-FGSM | ADS-QDir |

**Figure 6.** Qualitative comparison for Pulsar carriers at $256 \times 256$. Columns show the input carrier, JPEG $Q = 70$, diffusion 1-step, ADS-FGSM ($t = 1, \epsilon_{adv} = 0.01$), ADS-QDir ($t = 1, \epsilon_{adv} = 0.01$). ADS preserves perceptual content while substantially disrupting decoder recovery.

We next sweep the adversarial step size $\epsilon_{adv}$, since it is the central control that trades decoder disruption for perceptual distortion. Table 4 and Fig. 7 illustrate the security–utility trade-off as $\epsilon_{adv}$ increases. For ADS-FGSM, the trend is monotonic mainly over the tested range: $\epsilon_{adv} = 0.005$ already achieves 98.4% failure at LPIPS = 0.0021, and larger $\epsilon_{adv}$ values saturate failure at 100% while further increasing LPIPS.

## 4.4 Security–utility ablation study

**Table 4.** ADS sweep ($t = 1$) and corresponding security–utility points.

| Variant | $\epsilon_{adv}$ | Failure rate(%) | PSNR | SSIM | LPIPS | Latency (ms) |
|---|---|---|---|---|---|---|
| Diffusion 1-step | 0.000 | 7.81 | 50.54 | 0.997 | 0.0008 | 27.72 |
| ADS-FGSM | 0.005 | 98.43 | 46.49 | 0.994 | 0.0021 | 83.38 |
| ADS-FGSM | 0.010 | 100.00 | 43.80 | 0.988 | 0.0048 | 82.75 |
| ADS-FGSM | 0.020 | 100.00 | 39.82 | 0.968 | 0.0202 | 82.97 |
| ADS-QDir | 0.005 | 40.62 | 48.05 | 0.996 | 0.0014 | 82.37 |
| ADS-QDir | 0.010 | 98.43 | 46.31 | 0.994 | 0.0022 | 82.86 |
| ADS-QDir | 0.020 | 100.00 | 43.60 | 0.988 | 0.0048 | 82.54 |

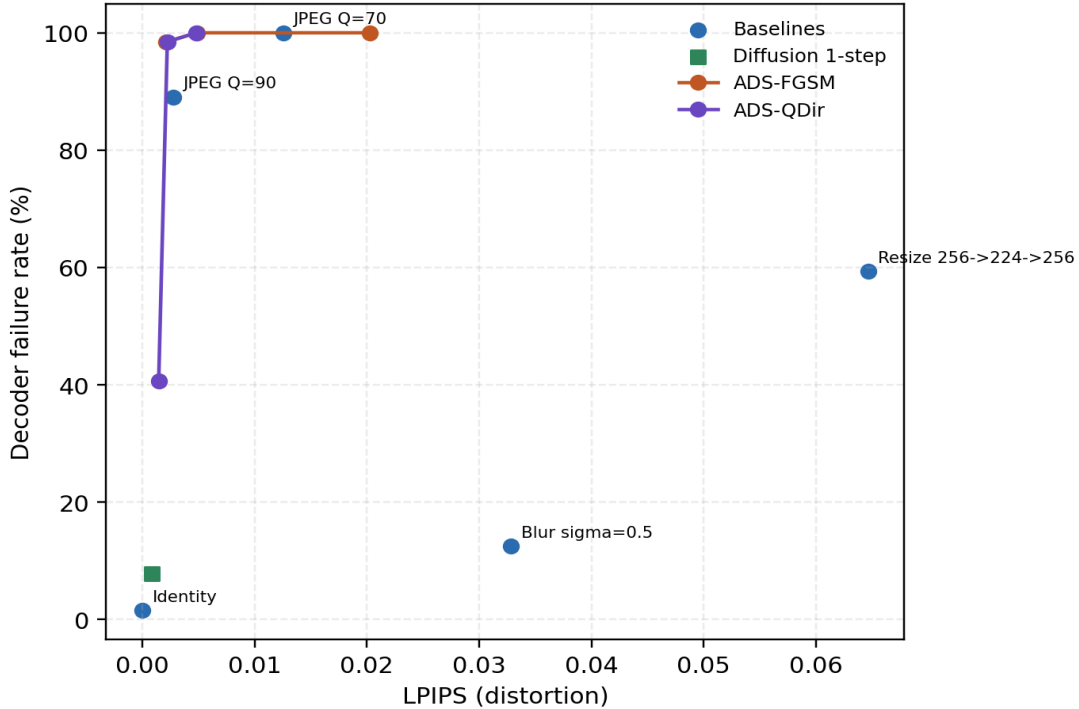

**Figure 7.** Decoder failure rate versus LPIPS distortion for Pulsar under gateway sanitization ($N = 64$). ADS variants trace a security–utility frontier as $\epsilon_{\text{adv}}$ increases at fixed $t = 1$. Conventional transforms (JPEG/resize/blur) appear as isolated operating points. ADS occupies the high-failure / low-distortion region, achieving near-saturated failure at LPIPS $\approx$ 0.002–0.005.

ADS-QDir shows a more gradual change at small step sizes. At $\epsilon_{adv} = 0.005$, it achieves 40.6% failure at LPIPS = 0.0014, and it reaches 98.4% failure at $\epsilon_{adv} = 0.01$ while remaining at LPIPS = 0.0022. This behavior matches the goal of QDir: by coupling RGB channels in the update direction, it avoids strong per-channel sign flips, enabling lower distortion when only partial disruption is desired, while still reaching near-saturated disruption once $\epsilon_{adv}$ is increased.

The frontier in Fig. 7 shows that ADS occupies the favorable region of high failure at low LPIPS compared to standard transforms. The most relevant comparison is at similar perceptual distortion (LPIPS $\approx$ 0.002–0.005). In this range, JPEG $Q = 90$ achieves 89.1% failure at LPIPS = 0.0028, whereas ADS-QDir achieves 98.4% failure at LPIPS = 0.0022, and ADS-FGSM achieves 100% failure at LPIPS = 0.0048. In contrast, mild blur and resize can introduce noticeably larger distortion without reliable disruption (e.g., blur $\sigma = 0.5$ yields only 12.5% failure at LPIPS = 0.0329). This indicates that higher distortion alone is not a reliable proxy for security against diffusion-stego decoders.

### 4.5 Latency considerations

ADS provides stronger disruption than classical transforms, but it increases the computational cost. On the measured GPU, JPEG/resize/blur requires about 1–2 ms per image, the diffusion 1-step baseline requires 27.7 ms, and ADS requires approximately 83 ms per image. The additional cost in ADS mainly comes from the backward pass and the extra proxy evaluation needed to compute the adversarial update. In deployment, this overhead can be reduced by batching and by applying ADS only at higher-risk ingress points. Overall, the measured latency remains below 100 ms per image on a single modern GPU, while providing a stronger security–utility trade-off compared to an ECC-enabled decoder.

# 5. Conclusions

This paper treats diffusion-based information hiding as a security gateway problem, where accurate and trustworthy detection cannot be assumed. To address this constraint, we adopt a sanitize-by-default strategy, treating every inbound image as untrusted. We introduce Adversarial Diffusion Sanitization (ADS), a training-free defense that applies a single forward diffusion step, performs a targeted adversarial update in diffusion space using a fixed pretrained denoiser as a differentiable proxy, and concludes with a single reverse step to produce a sanitized output. This design aims to disrupt decoder inversion stability while maintaining perceptual fidelity.

Our experiments with the Pulsar diffusion steganography baseline confirm the limitations of simple diffusion round trips: a near-identity forward and reverse step causes minimal disruption, resulting in a Decoder Success Rate (DSR) of 0.9219. Conversely, adding an adversarial diffusion-space update results in nearly complete decoder failure at low perceptual distortion. With $t = 1$ and $\varepsilon\_adv = 0.01$, ADS achieves 98.4%–100% decoding failure, while keeping LPIPS within 0.0022–0.0048. Moreover, the quaternion-coupled color update delivers comparable disruption with improved fidelity compared with uncoupled per-channel perturbations, achieving higher SSIM and lower LPIPS under identical distortion budgets.

Compared to standard gateway transforms that cause similar distortion, ADS significantly enhances the disruption of diffusion-stego decoding. JPEG recompression at quality 90 yields noticeably fewer failures at comparable LPIPS, whereas quality 70 recompression requires far higher distortion to disable decoding fully. These findings indicate that adversarial perturbations in diffusion space provide a more effective and distortion-efficient mechanism for interrupting diffusion-stego recovery than conventional content transformations.

Overall, diffusion-driven, coverless steganography poses a growing threat to content pipelines, enabling hidden communication channels that evade traditional detection methods. Adversarial Diffusion Sanitization (ADS) offers a practical and easily deployable solution: it leverages readily available denoisers as proxies and applies a color-sensitive, channel-coupled update. This approach effectively reduces decoder success while respecting perceptual constraints. By moving the focus from detection to proactive neutralization, ADS enhances the protection of synthetic-media ecosystems against diffusion-based steganographic risks.

Future work involves:

1. Developing adaptive proxy ensembles to enhance robustness across various sampling methods and diffusion schedulers.
2. Implementing content-aware distortion budgeting that focuses on regions most critical to decoder stability, while maintaining key semantics.

3. Extending capabilities to multimodal carriers such as video, audio, and network-layer payloads to expand protection throughout modern content workflows.
4. Developing an Adversarial Diffusion Sanitization (ADS), a training-free defense specifically designed for smartphones to combat stegomalware. Since smartphones are equipped with many sensors that gather data and hold substantial personal information, they are vulnerable to extensive profiling campaigns.